\documentclass[12pt,axodraw]{article}
\usepackage{epsf,epsfig}

\begin{document}

\begin{titlepage}

\begin{center}

\hfill ICRR-Report-514-2004-12\\
\hfill STUPP-05-178\\
\hfill KEK-TH-1013\\
\hfill TUM-HEP-584/05\\
\hfill \today

{\large 
Significant effects of second KK particles \\ on LKP dark matter physics
}
\vspace{1cm}

{\bf Mitsuru Kakizaki}$^{a,\,}$\footnote{kakizaki@icrr.u-tokyo.ac.jp},
{\bf Shigeki Matsumoto}$^{a,b,\,}$\footnote{smatsu@post.kek.jp},
{\bf Yoshio Sato}$^{c,d,\,}$\footnote{yoshio@krishna.th.phy.saitama-u.ac.jp} \\
and 
{\bf Masato Senami}$^{a,\,}$\footnote{senami@icrr.u-tokyo.ac.jp}
\vskip 0.15in
{\it
$^a${ICRR, University of Tokyo, Kashiwa 277-8582, Japan }\\
$^b${Theory Group, KEK, Oho 1-1, Tsukuba, Ibaraki 305-0801, Japan}\\
$^c${Department of Physics, Saitama University, Saitama 338-8570, Japan}\\
$^d${Physik-Department, Technische Universit\"at M\"unchen, \\ James-Franck-Strasse 85748 Garching, Germany}\\
}
\vskip 0.5in

\abstract{
We point out that Kaluza-Klein (KK) dark matter physics 
is drastically affected by second KK particles.
In this work 
various interesting phenomena caused by 
the second KK modes are discussed.
In particular,
we reevaluate the annihilation cross section 
relevant to the thermal relic density of KK dark matter
in universal extra dimension models.
In these models,
the first KK mode of $B$ boson is a viable dark matter candidate
by virtue of KK-parity.
We demonstrate that the KK dark matter annihilation cross section
can be enhanced,
compared with the tree level cross section mediated only by 
first KK particles.
The mass of the first KK mode of $B$ boson 
consistent with the WMAP observation is increased.
}

\end{center}
\end{titlepage}
\setcounter{footnote}{0}

% ------------
% Introduction
% ------------

\section{Introduction}

Precise measurements of the cosmological parameters 
have achieved amazing progress in recent years.
Especially observation of cosmic microwave background anisotropies
by WMAP revealed that non-baryonic dark matter amounts to 20 percent of 
the energy of our universe \cite{WMAP}.
The existence of the dark matter forces us to consider physics 
beyond the standard model (SM) for its constituent.
because there is no candidate for the dark matter in the SM.

On theoretical side 
weakly interacting massive particles (WIMPs)
are considered as excellent candidates for dark matter.
One of the excellent candidates is 
the lightest supersymmetric particle (LSP)
present in supersymmetric (SUSY) extensions of the SM.
The LSP is stabilized by virtue of R-parity and has been
most extensively studied so far \cite{reviews}.
Recently an alternative candidate for WIMP dark matter was proposed 
in universal extra dimension (UED) models \cite{LKP},
which is a well-motivated scenario invoking
TeV-scale extra dimensions \cite{TeVXD}.
In UED models, all SM particles propagate in the 
compact spatial extra dimensions,
leading to the towers of the Kaluza--Klein (KK)
partners for each SM particle in four dimensional point of view.
The KK mass spectra are quantized and each KK mode is
labeled by KK number $n$.
Momentum conservation in the compact extra dimension, 
which is converted to the KK number conservation, 
guarantees the stability of the lightest KK particle (LKP). 
The LKP is a viable candidate for dark matter.

The interesting aspect of UEDs is a characteristic feature of 
the mass spectra.
The masses of particles at each KK mode are highly degenerated,
because the compactification scale, 
the inverse of the size $R$ of extra dimensions, 
should be sufficiently large 
in order not to conflict with electroweak precision measurements
\cite{Appelquist:2000nn,EWmeasurement}.

In the simplest UED model,
the extra dimension is compactified on an $S^1/Z_2$ orbifold.
Orbifolding, required for obtaining chiral zero modes,
violates KK number conservation and leaves its remnant called KK-parity.
Under the parity 
particles at even (odd) KK modes have plus (minus) charge.
The parity also stabilizes the LKP.
This situation is quite similar to the LSP 
stabilized by R-parity in SUSY models.
The first KK mode of $B$ boson, $B^{(1)}$, 
is found to be the LKP in the minimal setup \cite{Cheng:2002iz}.

There have been several studies on this LKP,
which include the calculation of the relic density \cite{Servant:2002aq},
observational probabilities of dark matter through direct detections
\cite{Cheng:2002ej,direct} and 
indirect detections \cite{Cheng:2002ej},\cite{indirect}-\cite{positron},
and collider signatures at future accelerator experiments
\cite{Rizzo:2001sd}-\cite{collider}.
In particular its annihilation into fermion pair
is discussed extensively
because it has potential to account for the positron excess reported
by the HEAT experiment \cite{HEAT}.
It is difficult to explain the anomaly in the context of 
Majorana dark matter such as LSP.

% ------------
% Main subject
% ------------

In preceding studies contributions from first KK modes to various 
LKP phenomena have been extensively investigated,
whereas those from higher KK modes have not been considered.
In this paper we emphasize that the effects of second 
KK particles are important
and that it is worth investigating second KK physics in detail.
This is because dark matter is non-relativistic
from the age of the freeze-out to the present
and the incident energy of two LKPs
is almost degenerate with the masses of the second KK particles.
Thus the LKP pair annihilation processes mediated by second KK
particles in the $s$-channel may be significantly enhanced
due to the resonances.
Such resonances affect the relic abundance and
detection probabilities of dark matter in the present universe,
and collider phenomenology.
The situation should be contrasted with other models such as 
SUSY models, where tuning of fundamental parameters is required 
to achieve the resonance enhancement.
In this paper we appreciate the `natural resonance'
that incident energy of two first KK modes 
is close to masses of second KK particles.
This feature inherent in UEDs is applicable to many processes 
in LKP dark matter physics.

In particular we reexamine LKP abundance 
as one of the most important examples
in the second KK particle physics based on the minimal model.
We find that 
the $s$-channel dark matter annihilation process mediated by
the second KK mode of the neutral scalar Higgs boson, $h^{(2)}$,
competes with the tree level processes
in which no second KK particles contribute.
As a result the LKP mass consistent with 
the WMAP observation is increased
compared with that indicated in the preceding work \cite{Servant:2002aq}.

The outline of this paper is as follows:
in Sec. 2 we briefly review the minimal UED model based on
an $S^1/Z_2$ orbifold.
In Sec. 3 we evaluate LKP dark matter annihilation cross section.
We find the enhancement of the cross section due to
the resonance by the second KK Higgs boson $h^{(2)}$.
In Sec. 4 the relic abundance of the LKP dark matter is calculated.
The relic density of LKP turns out to be reduced by the enhanced 
cross section.
Other interesting phenomenology caused by the second KK particles
will be discussed in Sec. 5.
Section 6 contains our conclusions.

% --------------------
% Detailed explanation
% --------------------

\section{Universal extra dimension model}

We will briefly review the universal extra dimension model,
which is based on the hypothesis that
all SM particles propagate in the compact spatial extra dimensions
beyond the usual three dimensional space.
One of the attractive features of UEDs
is to provide a good candidate for cold dark matter.

At first we present the setup used in this letter.
Following the minimal UED model,
we postulate the field contents to be same as those of the SM.
We have three gauge fields $G$, $W$ and $B$, and one Higgs doublet $H$.
The matter contents are three generations of fermions:
the quark doublets $Q$, the up- and down-type quark singlets $U$ and $D$,
the lepton doublets $L$ and the charged lepton singlet $E$.
The extra dimension is compactified by an $S^1/Z_2$ orbifold.
Under the $Z_2$ parity
we assign plus charge to the left-handed doublets
and the right-handed singlets so that their zero modes
coincide with the SM chiral fermions.
The Higgs doublet must carry plus charge in order to allow the usual 
Yukawa couplings.

In the four dimensional perspective
the zero modes are identified with the SM particles
and their interactions are same as those in the SM.
The electroweak symmetry breaking also
mixes the hypercharge gauge boson $B$ and 
the neutral SU(2)$_L$ gauge boson $W^3$ at each KK level.
Since the mass difference between $B^{(n)}$ and $W^{(n)}$ $(n \geq 1)$
induced from radiative corrections is
larger than the electroweak scale,
we can neglect their mixing angles.
Therefore $B^{(n)}$ and $W^{(n)}$ are regarded as 
the mass eigenstates.

There are some possible candidates for LKP dark matter,
which must be neutral:
first KK states of 
the neutral gauge bosons, neutral Higgs bosons and neutrinos. 
The $W$ boson, the Higgs boson and the lepton doublet carry SU(2)$_L$
quantum number while the $B$ boson is singlet under the SM gauge group.
Thus radiative corrections increase the masses of 
$W^{(1)}$, $H^{(1)}$ and $L^{(1)}$, 
leaving the mass of $B^{(1)}$ almost unchanged \cite{Cheng:2002iz}.
As a result, $B^{(1)}$ is the LKP.

% Fig. 1
\begin{figure}[t]
  \begin{center}
    \scalebox{.5}{\includegraphics*{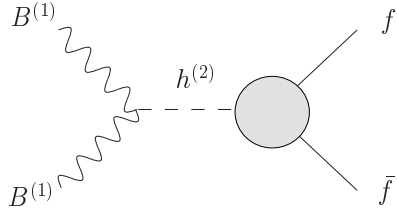}}
    \caption{{\footnotesize The resonant $B^{(1)}$ annihilation process 
      mediated by $s$-channel $h^{(2)}$ into a zero mode
      (SM) particle--anti-particle pair $f\bar{f}$.}}
    \label{fig:second}
  \end{center}
\end{figure}

\section{Annihilation cross section for LKP dark matter}
\label{sec:cross_section}

\begin{figure}[t]
  \begin{center}
    \scalebox{.5}{\includegraphics*{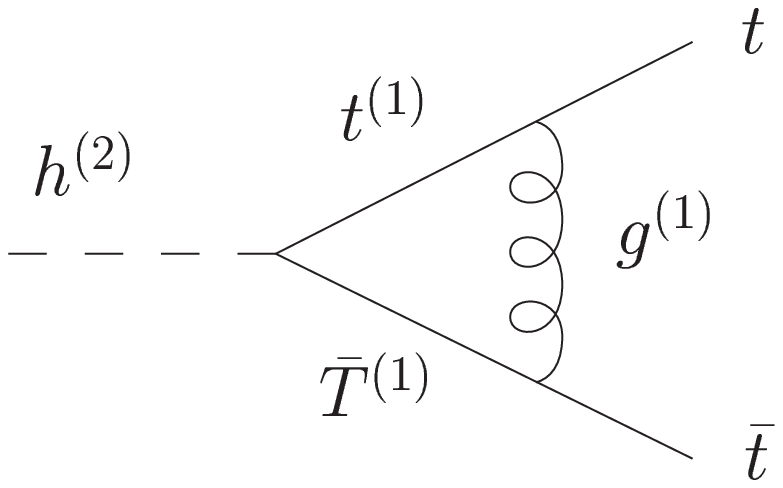}}
    \hspace{20mm}
    \scalebox{.5}{\includegraphics*{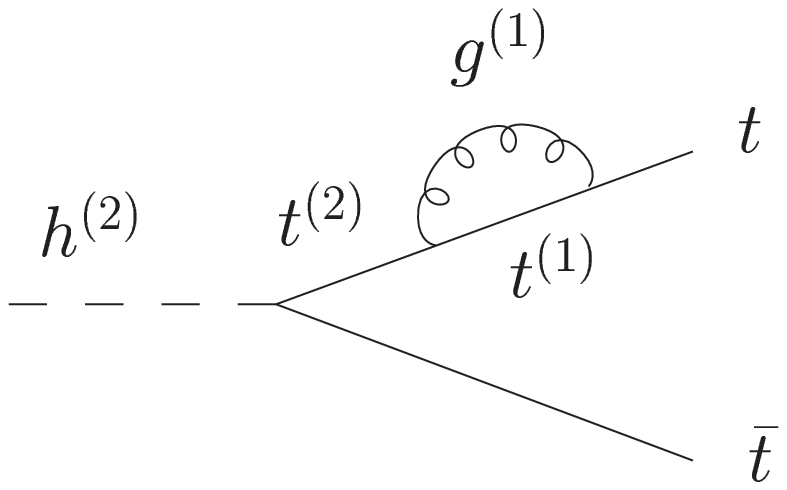}}
    \caption{{\footnotesize The dominant one-loop diagrams leading to 
      the resonant $B^{(1)}$ pair annihilation process mediated by
      $h^{(2)}$. 
      Here $t$ is the (zero-mode) top quark, 
      and $t^{(1)}, T^{(1)}$ and $g^{(1)}$ represent the first KK 
      modes of left- and right-handed top quarks and gluon respectively.
      }}
    \label{fig:triangle-gluon}
  \end{center}
\end{figure}

In this section we evaluate annihilation cross section 
for LKP dark matter.
We focus on resonant KK dark matter annihilation process into 
zero mode (SM) particles,
which proceeds with $s$-chanel $h^{(2)}$ as illustrated 
in Fig. \ref{fig:second}.
The $s$-chanel annihilation at tree level
into two first KK particles
and those into a second KK particle and an SM one are
kinematically forbidden or extremely suppressed by their small
Yukawa couplings.
On the other hand,
interactions among one second KK particle and two SM particles
are necessarily generated through radiative corrections
and thus dominantly contribute 
to annihilation processes.
We survey possible diagrams leading to such resonant annihilations
and find that the dominant contribution comes from
the one-loop diagrams depicted in Fig. \ref{fig:triangle-gluon}.

% Fig. 2

We have calculated the annihilation cross section
including the effect of the $h^{(2)}$ resonance: 
$\sigma = \sigma_{\rm res} + \sigma_{\rm tree}$.
The first term $\sigma_{\rm res}$ 
arises from the resonant one-loop diagrams as
\begin{eqnarray}
  \sigma_{\rm res} = 
  \frac{\pi \alpha_{\rm em} \tan^2 \theta_W m_Z^2}{9 m \beta}
  \frac{\Gamma_{h^{(2)}}}
  {(s-m^2_{h^{(2)}})^2+m_{h^{(2)}}^2 \Gamma_{h^{(2)}}^2}
  \left(3 + \frac{s(s-4m^2)}{4m^4} \right),
\end{eqnarray}
where $\alpha_{\rm em}$ and $\theta_W$ represent 
the fine-structure constant and weak-mixing angle,
$m$ and $m_{h^{(2)}}$ are the masses of $B^{(1)}$ and $h^{(2)}$
respectively,
$s$ is the center-of-mass energy squared and
$\beta^2 = 1 - 4m^2/s$.
The decay rate of $h^{(2)}$, dominated by the mode into 
the $t\bar{t}$ pair as discussed above, is written
\begin{eqnarray}
  \Gamma_{h^{(2)}} = \frac{y_t^2 \alpha_s^2 m_{h^{(2)}} }{384 \pi^3} 
  \left[ \ln \left( \frac{\Lambda^2}{\mu^2} \right) \right]^2,
\end{eqnarray}
where $\alpha_s$ and $y_t$ are the strong and
top Yukawa coupling constants respectively.
In the logarithm, $\Lambda$ denotes the cutoff scale of the theory
and $\mu$ represents the renormalization scale characterizing the typical
energy, which is set to be $\mu = m (\simeq R^{-1})$.
Here, we take the leading logarithmically divergent part into account
in the calculation of the cross section.
Resonant dark matter annihilation is naturally 
realized in the framework of UEDs 
because $s \simeq (2m)^2 \simeq m^2_{h^{(2)}}$.
On the other hand,
the last term $\sigma_{\rm tree}$ stems from the tree level
diagrams, in which only first KK particles are exchanged
\cite{Servant:2002aq}:
\begin{eqnarray}
  \sigma_{\rm tree} 
  = \frac{95 \pi \alpha_{\rm em}^2}{81 \cos^4 \theta_W}
  \frac{10(2m^2+s)\tanh^{-1}\beta -7s\beta}{s^2 \beta^2}
  + \frac{\pi \alpha^2_{\rm em}}{6 \cos^4 \theta_W s \beta}.
\end{eqnarray}
The interferential contribution between 
the tree-level diagram into $t \bar{t}$ and the one-loop diagrams 
is negligible
because it suffers from the chirality suppression of the top quark mass.

% -----------
% Calculation 
% -----------

\begin{figure}[t]
  \begin{center}
    \scalebox{.4}{\includegraphics*{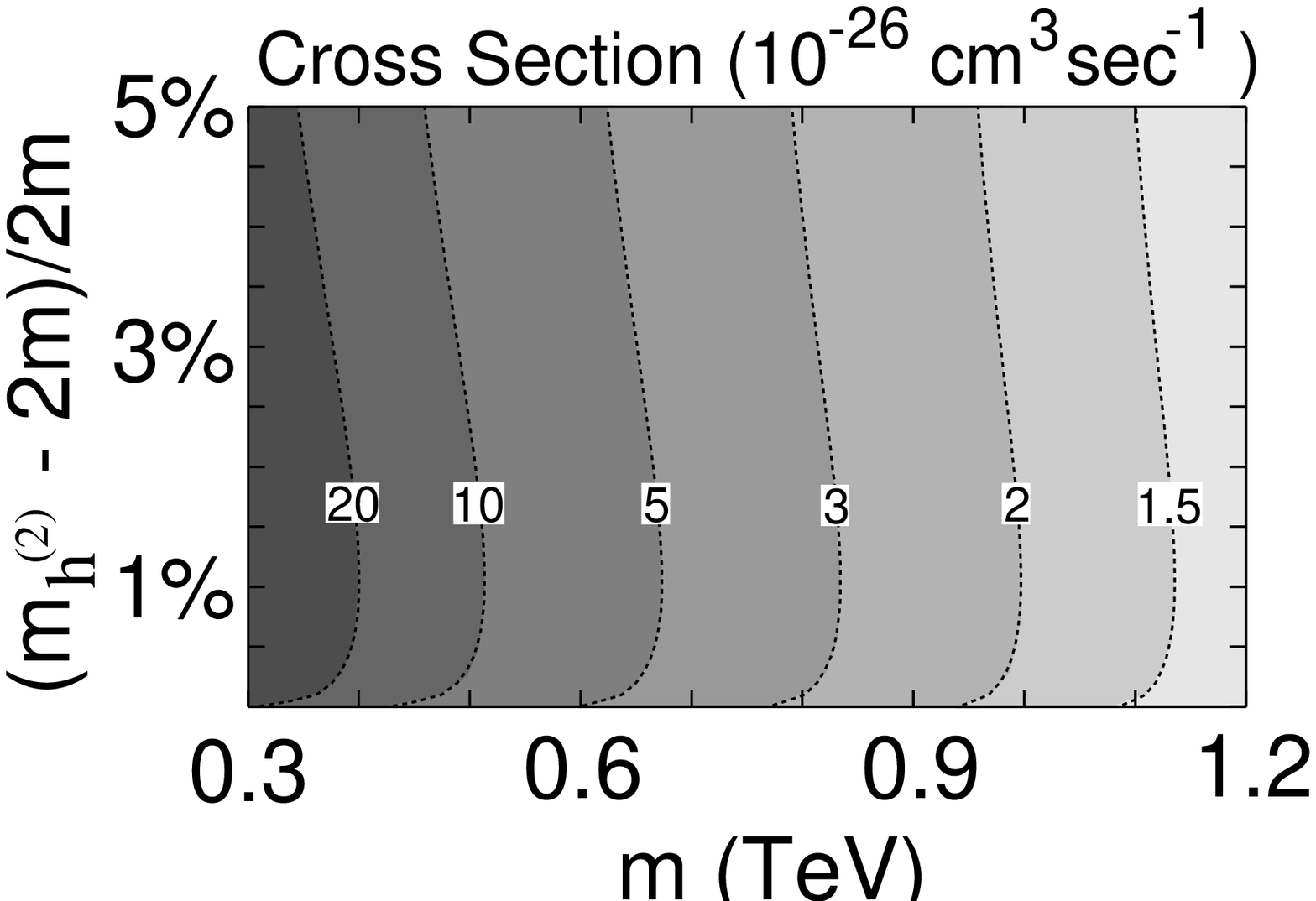}}
    \put(-100,-15){(a)}
    \hspace{0.5cm}
    \scalebox{.4}{\includegraphics*{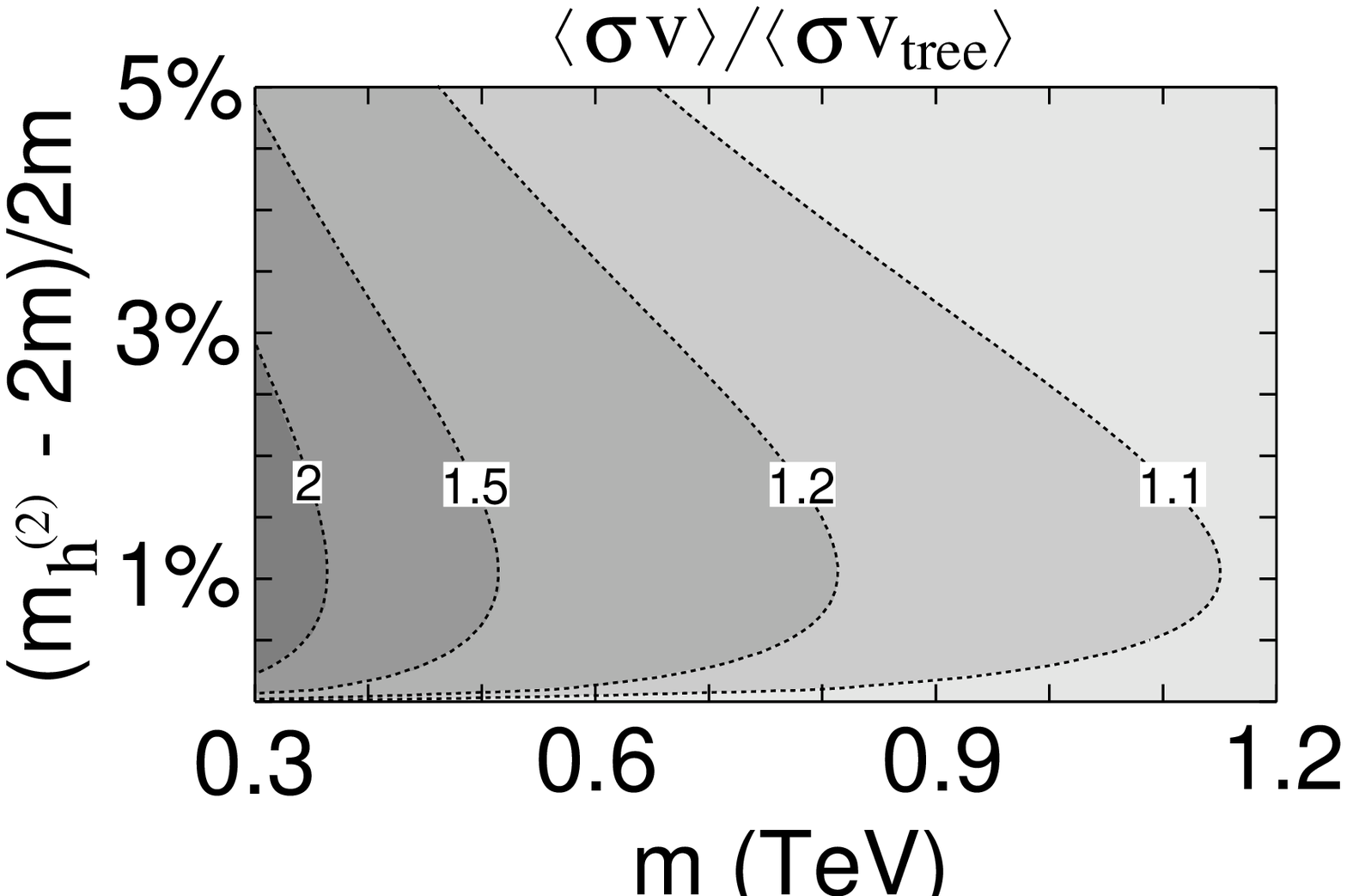}}
    \put(-100,-15){(b)}
    \caption{{\footnotesize (a) Contour plot 
        of the averaged annihilation cross section 
        (multiplied by the relative velocity) for the LKP, $B^{(1)}$. 
        (b) Contour plot of the ratio of the total averaged 
        cross section to the tree level one.
        The masses of $B^{(1)}$ and $h^{(2)}$ are represented by
        $m$ and $m_h^{(2)}$ respectively.
        Here we set the temperature to be $T=m/25$.
      }}
    \label{fig:CS}
  \end{center}
\end{figure}

We have numerically calculated
the thermally averaged annihilation cross section
(multiplied by
the relative velocity between incident $B^{(1)}$ bosons:
$v = \sqrt{s} \beta/m$) in the non-relativistic limit,
which is given by
\begin{eqnarray}
  \langle \sigma v \rangle = 4 \pi \left(\frac{m}{4 \pi T} \right)^{3/2}
  \int_0^\infty dv \ v^2 e^{- m v^2 / 4 T} \sigma v, 
\end{eqnarray}
at temperature $T$.
It is important to evaluate the thermal average of the cross section
at the freeze-out temperature because it affects the calculation of
the abundance of the thermal relic.
In case of the UED model the LKP decouples from thermal equilibrium 
at $T \sim m/25$.
Figure \ref{fig:CS}(a) shows 
the contour plots of the predicted annihilation cross section 
in units of $10^{-26}\ \mbox{cm}^3 \ \mbox{sec}^{-1}$.
The ratio of the total cross section to that at tree level
is also depicted in Fig. \ref{fig:CS}(b).
We choose the temperature as $T = m/25$. 
The cutoff scale is set to be $\Lambda R = 20$,
which is a characteristic value assumed in the UED scenarios.
Here we treat $m_{h^{(2)}}$ as a free parameter instead of the 
Higgs mass and show
the dependence of the cross sections on the mass splitting 
$\delta \equiv (m_{h^{(2)}}-2m)/2m$.
These figures demonstrate that 
the resonant process competes with those at tree level,
resulting in increased annihilation cross section.
For $\delta \sim 1\ \% $, the incident energy of the LKP pair
matches the $h^{(2)}$ pole and leads to enhancement.
It is interesting to notice that 
$\delta \simeq 1\ \% $ is indeed realized in a
wide region of the parameter space 
in the minimal UED \cite{Cheng:2002iz}.

% Fig. 3

\section{Thermal relic density of LKP dark matter}
\label{sec:relic}

We now calculate the thermal relic density of the LKP dark matter
using annihilation cross section obtained 
in Sec. \ref{sec:cross_section}.
The time evolution of the LKP number density $n$
obeys the following Boltzmann equation:
\begin{eqnarray}
  \frac{dn}{dt} + 3 H n 
  = - \langle \sigma v \rangle (n^2 - n_{\rm eq}^2 ),
\end{eqnarray}
where $H$ is the Hubble parameter describing 
the expansion of the universe.
In the non-relativistic limit
the equilibrium number density $n_{\rm eq}$ at temperature $T$
is given by 
\begin{eqnarray}
  n_{\rm eq} = g \left( \frac{m T}{2 \pi} \right)^{3/2} e^{- m / T},
\end{eqnarray}
where the number of degree of freedom is $g=3$ for the LKP.
Due to the conservation of entropy per comoving volume,
the Boltzmann equation is rewritten
\begin{eqnarray}
  \frac{dY}{dx} = 
  - \frac{\langle \sigma v \rangle}{Hx} s ( Y^2 - Y_{\rm eq}^2)
  \label{eq:BE}
\end{eqnarray}
where $Y = n/s$ and $Y_{\rm eq} = n_{\rm eq}/s$ and $x = m/T$
with entropy $s = 2 \pi^2 g_* T^3 / 45$.
Here $g_*$ denotes the number of relativistic degrees of freedom.
By solving the Boltzmann equation we obtain the present
number density over entropy of dark matter $Y_\infty$.
It is useful to express the relic density
in terms of $\Omega h^2$.
Here $\Omega$ is the ratio of the dark matter energy density
to the critical density in the present universe.
The dark matter density is given 
by $\rho = m n = m s_0 Y_\infty$, with $s_0$ being 
the present entropy $s_0 = 2900 \ \mbox{cm}^{-3}$.
The critical density is
$\rho_c = 3 H_0^2 M_{\rm Pl}^2 / 8 \pi 
= 1.1 \times 10^{-5} \ h^2 \ \mbox{GeV} \ \mbox{cm}^{-3}$,
where the Planck mass is 
$M_{\rm Pl} = 1.2 \times 10^{19} \ \mbox{GeV}$ 
and $H_0$ is the present Hubble expansion rate parameterized as
$H_0 = 100\ h \ \mbox{km} \ \mbox{sec}^{-1}\ \mbox{Mpc}^{-1}$.
The WMAP data is fitted by $h = 0.72 \pm 0.05$ \cite{WMAP}.

% Fig. 4
\begin{figure}[t]
  \begin{center}
    \scalebox{.3}{\includegraphics*{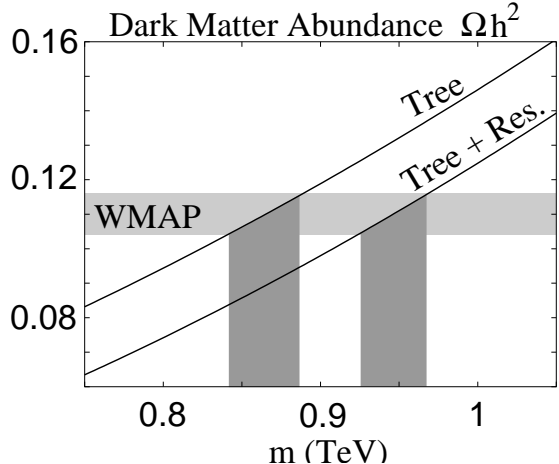}}
    \caption{{\footnotesize 
        The predicted dark matter abundance $\Omega h^2$ as a
        function of the LKP mass $m$ including resonance (Tree + Res.)
        for $g_* = 100$.
        Here we set the cutoff scale to be $\Lambda R = 20$ and 
        the mass splitting $\delta = 1\ \%$.
        For comparison, we show the tree level result (Tree).
        The $1 \sigma$ region of the relic abundance measured by WMAP 
        is also shown: $\Omega h^2 = 0.110 \pm 0.006$.
        The allowed mass regions are highlightened in both cases.}}
    \label{fig:Omegah2}
  \end{center}
\end{figure}

We numerically solve the Boltzmann equation in Eq. (\ref{eq:BE}) to
obtain the thermal relic density.
Figure \ref{fig:Omegah2} shows
the predicted dark matter abundance $\Omega h^2$ (Tree + Res.)
as a function of the LKP mass $m$ for $g_* = 100$.
Here we take $\Lambda R = 20, \ \delta = 1\ \%$ for UED parameters.
For comparison, we show the tree-level result
(Tree) \cite{Servant:2002aq}.
The $1 \sigma$ region of the relic abundance measured by WMAP
is also shown: $\Omega h^2 = 0.110 \pm 0.006$.
The mass regions accounting for the dark matter density
are highlightened in both cases.
As a consequence of the resonant annihilation,
the mass of the KK dark matter
consistent with the WMAP data turns out to be around 
$950 \ \mbox{GeV}$, which is
$\sim 100\ \mbox{GeV}$ above the tree-level result.
The relic abundance depends on the mass splitting $\delta$,
although the numerical difference is small:
the allowed region is found to be $m \simeq 900 - 1000$ GeV 
for $0.5 \ \% < \delta < 2 \ \%$.
We have also checked that 
the relic abundance is almost insensitive to the cutoff scale $\Lambda$.
We conclude that the reduction of the LKP dark matter abundance
due to the second KK resonance is a characteristic feature of UED 
models and independent of the detailed mass spectra.

% -------------------------
% Discussion and conclusion
% -------------------------

\section{Phenomena affected by second KK particles}
\label{sec:discussion}

% Fig. 5
\begin{figure}[t]
  \begin{center}
    \scalebox{.36}{\includegraphics*{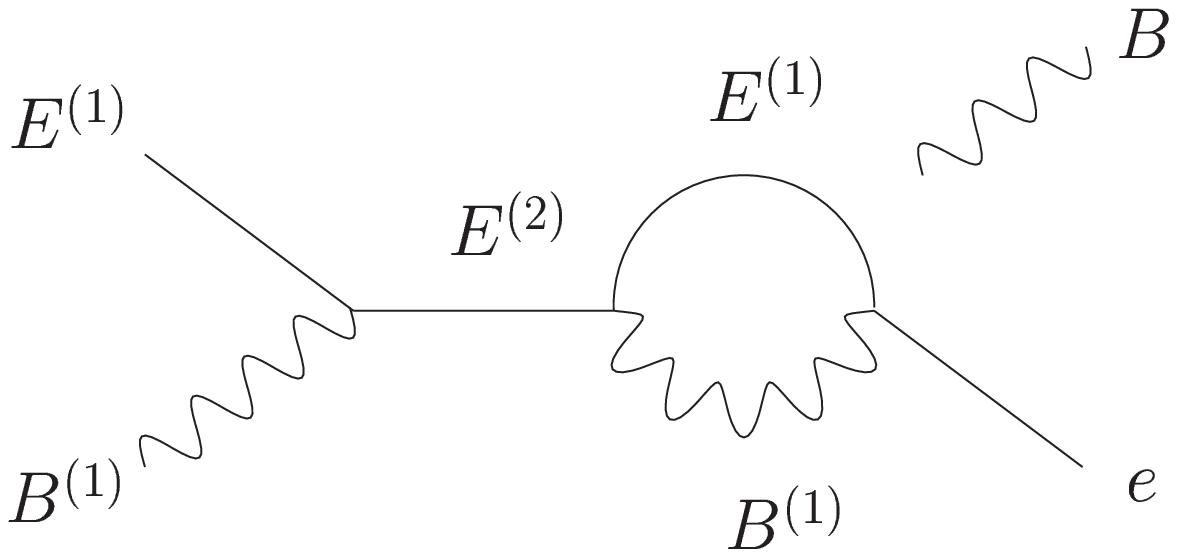}} 
    \put(-70,-15){(a)}
    \hspace{5mm}
    \scalebox{.36}{\includegraphics*[bb = 118 640 482 769]{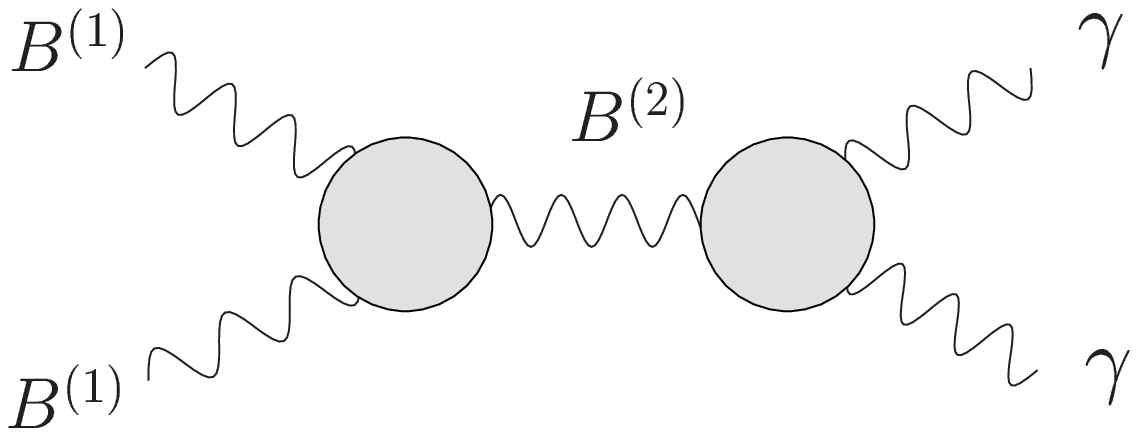}}
    \put(-70,-15){(b)}
    \hspace{5mm}
    \scalebox{.36}{\includegraphics*{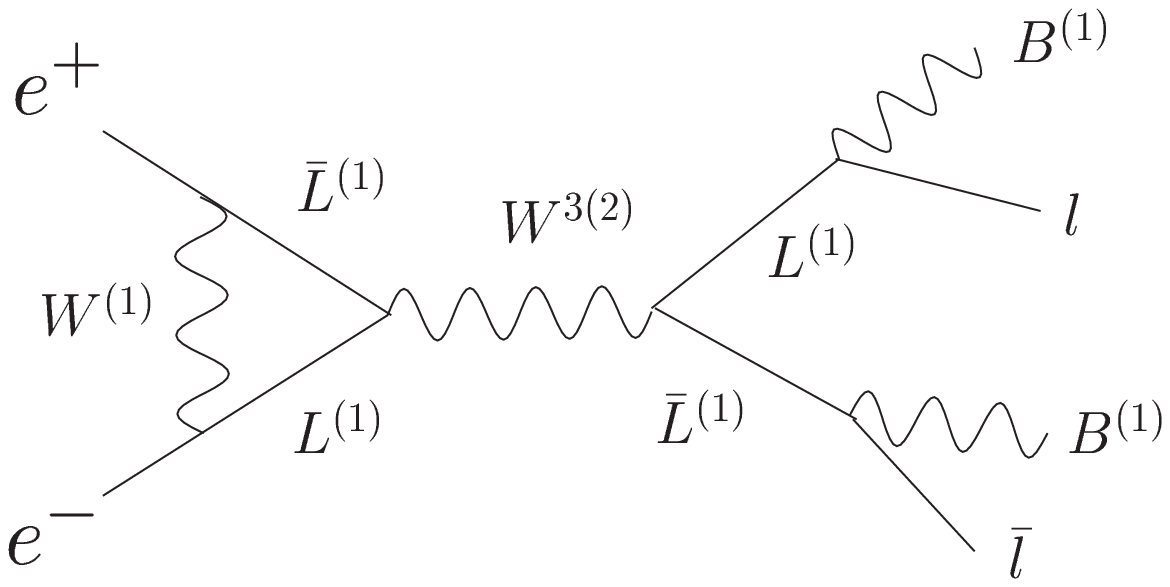}}
    \put(-70,-15){(c)}
    \caption{{\footnotesize Resonant processes induced by second KK particles
      which influence (a) coannihilation, (b) indirect detection and
      (c) collider phenomenology.
      In the diagrams $e,l$ and $\gamma$ denote the zero mode (SM) 
      electron, charged lepton and photon,
      $E^{(1)}, L^{(1)}$ and $B^{(1)}$ represent
      the first KK modes of right- and left-handed charged lepton,
      $B$ and Higgs boson, and
      $E^{(2)}, B^{(2)}$ and $W^{(2)}$ stand for the second KK modes
      of the right-handed charged lepton, $B$ and $W$ bosons.
    }}
    \label{fig:res_diagrams}
  \end{center}
\end{figure}

In Sec. \ref{sec:relic} we observed that 
the predicted relic density of LKP dark matter is
significantly changed by the resonance caused by the intermediate 
second KK Higgs boson compared with the tree level result.
Here we discuss other
phenomenological issues originating from second KK physics.
The `natural resonance' is a generic feature
appearing in UEDs and applicable to various processes of LKP dark matter
physics.
In the following we consider the cases where
the second KK physics plays an essential role
in making exciting predictions:

{\it Coannihilation:}
The predicted relic density of dark matter
is drastically changed by the so-called coannihilation 
when other particles are highly degenerate with the LKP in mass.
For the tree level calculation of the relic density,
it is known that the LKP mass consistent with WMAP decreases
in the presence of the degenerate 
first KK right-handed electron $E^{(1)}$ \cite{Servant:2002aq}.
The self-annihilation rates of $B^{(1)}$ and $E^{(1)}$
are of the same order while the coannihilation rate is rather suppressed.
Then after decoupling from thermal equilibrium,
more relics are left over, 
which lowers the allowed LKP mass.
However 
the coannihilation of $B^{(1)}$ with $E^{(1)}$ is also affected by
`natural resonance' through $s$-channel $E^{(2)}$ exchange
accompanied by dipole-type interaction depicted in Fig. 
\ref{fig:res_diagrams}(a).
The coannihilation rate may be
comparable to the self-annihilation rates under certain circumstances,
and the allowed LKP mass is also modified 
from that calculated at tree level by coannihilation.
The detailed analysis including the coannihilation effects
will be presented in the subsequent paper \cite{KMSS}.

{\it Indirect detection:}
Since dark matter is almost at rest in the present universe,
the energy of two LKPs matches masses of second KK modes.
Especially the relation $m_{B^{(2)}} = 2 m_{B^{(1)}}$ is almost retained
even after inclusion of radiative corrections to mass spectra.
Thus, large annihilation cross sections are induced by
the two-loop processes mediated by $B^{(2)}$ in the $s$-channel
as depicted in Fig \ref{fig:res_diagrams}(b).
The monochromatic gamma-rays are produced by the LKP annihilations
\cite{Bergstrom:2004nr}.
We find that the resonant $B^{(2)}$ process into two photons 
at the two-loop level
gives a contribution comparable to the one-loop processes 
mediated by (KK) fermions.
The LKP annihilations also directly yield a large number of positrons 
whose energy is equal to the LKP mass, in contrast to the LSP case.
The contribution from the $B^{(2)}$ resonance 
is found to change the predicted positron flux by ${\cal O}(10)$ percent.
We leave these interesting subjects to future works.

{\it Collider signatures:}
Future accelerator experiments are very promising to probe UEDs.
In the simple $S^1/Z_2$ case
the discovery reach at the Large Hadron Collider will extend to 
$R^{-1}\sim 1.5 \ \mbox{TeV}$ \cite{Cheng:2002ab},
well above the compactification 
scale favored by the LKP dark matter scenario.
Furthermore, lepton colliders with sufficient 
center-of-mass energy could conclusively test the UED model
due to the `natural resonance'.
Let us consider pair production of 
the first KK left-handed charged lepton
mediated by the $s$-channel second KK mode of $W$ boson
as shown in Fig. \ref{fig:res_diagrams}(c).
Each product decays into a charged lepton and 
the LKP carrying large missing energy.
The signal event of two charged leptons plus missing energy
is expected with significantly large cross section
and almost background-free \cite{Battaglia:2005zf}.
Since the degeneracy between the intermediate particle mass and 
twice the LKP mass 
is a characteristic feature in the framework of UEDs,
we could discriminate UEDs from other new physics.

\section{Conclusion}
\label{sec:conclusion}

We have investigated
the effects from the second KK particles
on low energy predictions.
We especially focused on the relic abundance of LKP dark matter, which
is the first KK mode of $B$ boson, $B^{(1)}$.
We have pointed out that
due to the $h^{(2)}$ resonance
the dark matter annihilation cross section is enhanced compared to
that involving only the first KK modes.
As a result the allowed value of $B^{(1)}$ mass constrained by the 
WMAP observation is shifted to a larger mass region.
We have also discussed
various interesting phenomena caused by the second KK 
modes including coannihilation, indirect detection and 
collider signatures.
Studies involving `natural resonance' will make tremendous advance 
in KK particle physics.

\section*{Acknowledgments}
The work of M.K. is supported in part by
the Japan Society for the Promotion of Science.

% ----------
% References
% ----------

\end{document}